# Evaluation of an Intelligent Assistive Technology for Voice Navigation of Spreadsheets

Derek Flood, Kevin Mc Daid, Fergal Mc Caffery, Brian Bishop
Dundalk Institute of Technology,
Derek.flood@dkit.ie, Kevin.mcdaid@dkit.ie, Fergal.McCaffery@dkit.ie,
Brian.bishop@dkit.ie,

## ABSTRACT

*An integral part of spreadsheet auditing is navigation. For sufferers of Repetitive Strain Injury who need to use voice recognition technology this navigation can be highly problematic. To counter this the authors have developed an intelligent voice navigation system, iVoice, which replicates common spreadsheet auditing behaviours through simple voice commands. This paper outlines the iVoice system and summarizes the results of a study to evaluate iVoice when compared to a leading voice recognition technology.*

## 1       INTRODUCTION

Two independent studies [Panko, 2005, Powell, 2007] have shown that over 90% of operational spreadsheets contain errors. These errors can cause a spreadsheet's bottom line values to be off by thousands of dollars. One such error in a spreadsheet, caused the University of Toledo to introduce sudden budget cuts amounting to $2.4 Million (http//www.eusprig.org/stories.html).

Many approaches, some of which are outlined in Section 2, have been proposed to help improve the quality of spreadsheets. These approaches fall into two broad categories. The first category attempts to prevent errors from entering a spreadsheet through improved development methodologies. Test First Development [Mc Daid, 2008] is one such approach. The second category accepts that errors will always be present in spreadsheets and focuses on locating and removing these errors. Spreadsheet debugging falls into the second category.

Spreadsheet debugging is a process by which an individual reviews a spreadsheet for errors and removes those that are found. For users of voice recognition technology, such as those suffering from RSI (Repetitive Strain Injury), this process can be difficult due to the large number of commands that are required and the recognition times associated with these commands. A recent study [Flood, 2007] has shown that auditing spreadsheets through voice recognition technology is more time consuming and less accurate than conventional means.

Some software application domains, as outlined in Section 3, have been adapted to take advantage of the unique features offered by voice recognition. With the aim of improving the control of spreadsheets through speech, the authors have developed a tool called iVoice. By providing simple voice commands, based on the context of the currently selected cell, iVoice aims to simplify spreadsheet navigation. Following an introduction to spreadsheets and voice-control of software applications, Section 4 details the iVoice system.

A comparative study, as outlined in Section 5, was conducted to evaluate iVoice compared with the current state of the art voice recognition technology. Section 6 summarizes the results of this study. Potential improvements suggested by study participants are outlined in Section 7. Section 8 concludes this paper.

## 2  SPREADSHEET TECHNOLOGY

Spreadsheets are a key technology in modern business. Within the financial district of London, Spreadsheets have been described as "the primary front line tool of analysis" [Croll, 2005]. Although the decisions being made on the information contained in spreadsheets are of great importance, there is very little done to ensure the quality of these spreadsheets.

Two independent studies [Powell, 2007, Panko, 2005] have revealed that over 90% of spreadsheets contain errors. These errors can range in severity from simple spelling errors to complicated formula errors, which can have serious financial consequences. There have been many reported cases of such consequences including the Columbia Housing Authority who overpaid landlords by $118,387 as a result of a spreadsheet error [Miller, 2006].

A range of techniques have been proposed to improve the quality of spreadsheets. Some of these initiatives focus on the importance of development methodologies similar to existing lifecycle processes in place for software development.

Test Driven Development of spreadsheets [Rust, 2006, Mc Daid, 2008] is one such example. This method, which is growing in popularity in software development, requires users to write test cases before they develop the spreadsheet. In this way developers will have a series of tests in place that will ensure that the formulas that are implemented are correct.

Other approaches, such as What You See Is What You Test (WYSIWYT) [Fisher, 2006, Burnett, 2002], try to identify the location of faults once the spreadsheet has been created. WYSIWYT is used for fault localization. Users test a cell and notify the system, through a checkbox, whether the value is right or wrong. If the value is wrong the system can determine, based on measures adapted from software testing coverage techniques, the likely causes of the problem.

## 3  VOICE RECOGNITION TECHNOLOGY

Voice recognition is a process by which audio input is translated into textual information. Voice command technology, or voice interfaces, use voice recognition to allow a user to operate a computer system with their voice. Voice recognition is not new, however with the increase in computer power in recent years this technology has improved to the point that accuracy of voice recognition tools is now advertised to be of the order of 99%.

Voice recognition technology has been successfully used for creating and modifying text documents [Scansoft, 2004]. Within this domain it has been demonstrated to be faster than the traditional keyboard & mouse input. Through the use of the voice recognition software Dragon NaturallySpeaking, the Social Security Administration in the US, who deal with incorrectly paid benefits, were able to reduce the turnaround time for appeals from 44 days to 22 days, saving tax payers $500,000 [Scansoft, 2004]. This was achieved by allowing judges, who review these cases, to dictate their findings through Dragon NaturallySpeaking (DrNS).

With the increased interest in voice technology, new applications across a range of domains have emerged. Begel [Begel, 2004], through the *Spoken Java* technology, has looked at using a voice interface to allow developers to create Java programs. *Spoken Java* is a verbalizable alternative to the Java programming language, which allows developers to speak code in a natural way, having the system generate the Java source code.

Other domains such as 3-D animation have also successfully adopted the use of voice technology. Wang [Wang, 2006] developed a 3-D animation system that allows an animator to use voice recognition technology to create animations. The animator can issue commands to objects and have them perform a predefined action, much the same way as a traditional director would tell actors/actresses what they need them to do.

There are many voice recognition engines available; the best selling of these is Dragon NaturallySpeaking. Dragon NaturallySpeaking boasts an accuracy of 99%, which means it will correctly recognise 99 out of every 100 words. The most basic version, *Standard Edition*, features support for Microsoft© Word and Microsoft© Internet Explorer. The *Preferred Edition* also includes support for Microsoft© Excel, but there is no indication as to its efficiency in this domain. Special editions for legal and medical domains also exist .

## 4      IVOICE

In an experiment described in [Flood, 2007] the authors found that spreadsheet auditing through voice recognition was slower and less accurate than traditional keyboard & mouse control.

The experiment asked three experienced spreadsheet users to audit a spreadsheet seeded with errors using a state of the art voice recognition technology, Dragon NaturallySpeaking. For the experiment, the Preferred edition of Dragon NaturallySpeaking was used as it facilitates the use of voice recognition within Microsoft© Excel, which was used as the spreadsheet application. None of the participants had any prior experience with voice recognition, however they had sufficient experience with spreadsheets to complete the task easily. Their performance was compared with another study [Bishop, 2007] in which Bishop asked 13 professional spreadsheet users to audit the same spreadsheet using a keyboard and mouse. The study recorded the behaviour of each participant through logging of cell selection activity via the T-CAT [Bishop, 2008] (Time-stamped Cell Activity Tracker) tool. This technology was also used for the analysis of the voice-controlled subjects. Overall it was found that, despite taking twice as long as the keyboard and mouse participants, those using voice recognition found 14% less errors.

A number of actions including editing formulas, entering data and navigation were examined to identify key differences in the behaviour and performance of the groups. The results showed that the voice control users struggled in all these aspects of the task. Given that navigation is the most fundamental of spreadsheet activities for spreadsheet testers, it was decided to first explore the development of technologies that would improve the efficiency of voice-controlled navigation of spreadsheets.

The resulting iVoice system, which integrates with Dragon NaturallySpeaking (DrNS), provides support for three particular actions, namely navigation to referenced cells, automatic scanning of a range of cells and navigation directly to the next non-blank cell. A number of other elements of navigation were also looked at such as navigation within a formula and direct navigation to a worksheet. These may be the subject of future work. Each of the three features are now explained and subsequently investigated through a controlled study.

## 4.1    NAVIGATION TO REFERENCED CELLS

When looking at the navigational behaviour of both keyboard/mouse and voice-control participants, it was noticed that on initially entering a cell with a reference to a remote cell users tended to go to that remote cell to ensure that the reference was correct. However in the case of linked cells located in separate worksheets, traditional voice recognition software requires users to navigate to the target worksheet through all intermediary worksheets, as Dragon NaturallySpeaking only provides two commands for changing worksheet, "*Next*

*Worksheet"* and *"Previous Worksheet"*. The order of the worksheets is determined by the order of the worksheet tabs. iVoice allows users to navigate directly to the referenced worksheet, thus bypassing all intermediate worksheets.

By analysing the formula in the current cell and assigning each referenced cell a unique colour, users can, through the command "*Jump <colour>*" where *<colour>* is the colour of the desired destination cell, move directly to the required cell. A "*Jump Back*" command is also provided to enable users to move back to the original cell. It is hypothesised that the time to reach a referenced sheet will, based upon the number of commands, decrease for non adjacent worksheets using the iVoice technology.

Figure 1: Navigation to Referenced Cell

If the referenced cell is visible on screen iVoice will set the background colour of that cell. In Figure 1, the cell E6 can be seen with a green background, therefore to move to E6 a user would only need to say "*Jump Green*". If a cell is not visible on screen the system displays a coloured box in the lower right hand side of the screen. In Figure 1, the chosen cell references cell D6 on the *Opening Stock* worksheet, and iVoice has added a pink box in the lower right hand corner of the screen to indicate this reference has been assigned the colour pink. If the user says "*Jump Pink*" the focus will shift from the current cell to cell D6 on the *Opening Stock* worksheet.

The list of the colours and the names used by the system can be displayed, as in Figure 2, through the "*Show Colours*" command and hidden again with the "*Hide colours*" command. Currently the system uses only seven colours for referencing cells. If a cell contains more references than this, the system offers shortcuts to the first seven references only. When a range is encountered the system colours the first cell in the range to allow direct navigation to this cell. The colours used are based on those used by Microsoft© Excel for colouring references in edit-cell mode. It is hoped in the future to allow users define their own colours and colour names for use within the iVoice system.

Figure 2: iVoice Colours Help

### 4.2  SCAN REGION

The second component of the iVoice system allows users to navigate through a list of semantically similar cells quickly and easily. Semantically similar cells are those whose contents are of a similar structure and purpose. Spreadsheets are in general composed of regions of such semantically similar cells. It has been observed in a previous study [Flood, 2007], that users will examine these regions sequentially, spending on average between 0.33 seconds and 1.5 seconds on each cell.

iVoice proposes to support this navigation activity through the provision of a scan command which automatically moves to the next cell in the chosen direction after one second. This delay allows users the chance to review the contents of the cell before moving on. To initiate this command users say "*Scan <direction>*" where *<direction>* is the direction they wish to scan, be it "left", "right", "up" or "down". It is hoped that this feature will reduce the time to perform this task as it requires a single voice command rather than a series of such commands.

### 4.3  JUMP BLANK CELLS

The third component of the iVoice system allows users to skip over blank cells, by saying "*Jump <direction>*" where *<direction>* represents the way the user wants to move. The system moves directly to the next non blank cell in that direction. It is felt that this command is more efficient and natural than dictating the associated keyboard shortcut as is currently required in Dragon NaturallySpeaking.

## 5  COMPARITIVE STUDY

A controlled study was designed to compare the efficiency and effectiveness of the iVoice system with Dragon NaturallySpeaking (DrNS). This included a qualitative study, where participants took part in a structured interview to establish their view of the iVoice technology.

A quantitative experiment asked six experienced spreadsheet users to locate as many errors as they could in two spreadsheets through voice recognition technology. To randomise the experiment participants were split into two groups of three with Group 1 first auditing spreadsheet 1 using DrNS and then Spreadsheet 2 using iVoice. Group 2 reversed the use of the technologies, using iVoice for Spreadsheet 1 and DrNS on Spreadsheet 2. The cell selection behaviour of all participants was recorded through the T-CAT tool.

Each of the two spreadsheets were seeded with errors and participants were asked to highlight, by saying "Mark Error", any cells they believed to contain an error. If they later changed their mind they could say "Unmark Error" to remove the highlighting from that cell. The spreadsheets, while distinct, were similar in size and complexity.

Before the trial commenced users were asked to configure the voice recognition software to their voice. By reading aloud a passage of text the voice recognition software became more familiar with the participant's speaking pattern and thus improve the accuracy of the voice recognition software. This took approximately ten minutes to complete, after which users were introduced to the first voice-control system through a navigation exercise where they were required to navigate a sample spreadsheet.

When users felt they had mastered the navigation commands, the trial commenced with users auditing Spreadsheet 1. No time limit was set for the task allowing users to finish when they believed they could not find any more errors.

The first spreadsheet audited comprised three worksheets, *Wages*, *Expenses*, and *2007 Department Spending*. The spreadsheet calculated the total expenditure for each of three departments in a company. The *Wages* sheet detailed all employees wages and which department they belong to. The *Expenses* worksheet was used to detail different expenses to the company and what department these expenses should be assigned to. The final worksheet totalled all costs and apportioned company-wide expenses to each department based on the number of employees in that department.

There were 12 errors placed in Spreadsheet 1 some of which were formula and some data errors. During the trial participants were given an explanation of the spreadsheet, including a set of rules specifying the function and detailed design of the spreadsheet. When the first spreadsheet had been completed users were given a break to facilitate the changing of the technology. Users then repeated the training with the second technology and audited the second spreadsheet. Again, no time limit was set and users finished when they felt they could find no more errors.

The second spreadsheet again comprised of three worksheets, *Opening Stock*, *Purchases* and *Sales and Profit*. This spreadsheet was used to calculate the profit made on each of 18 products over a given period. The *Opening Stock* worksheet detailed the quantity and value of each of the products at the start of the period. The *Purchases* worksheet detailed the purchases that were made during the period, and the *Sales and Profit* worksheet detailed the sales and closing stock of each product. The *Sales and Profit* worksheet also uses the costs from the *Opening Stock* and *Purchases* worksheets to calculate the profit for the period. In Spreadsheet 2 there were 18 errors. Again participants were given a detailed specification for the spreadsheet.

## 6  ANALYSIS OF RESULTS

Before examining the individual features, we examine the overall performance and behaviour of participants using each technology. The T-CAT tool allowed for a detailed analysis of the experiment.

The measures used for performance are spreadsheet coverage, errors found and time taken. Bishop [Bishop, 2007] found evidence to suggest that there is a relationship between the number of cells evaluated and the number of errors found. Coverage based approaches to software testing are based on this belief.

Coverage was defined as the percentage of spreadsheet cells reviewed by participants, where a cell was considered to be reviewed if a participant spent more than 0.3 seconds in that cell. Only cells that contain numerical data or a formula were considered as other cells could be reviewed without being entered.

|  | Spreadsheet 1 | Spreadsheet 2 |
|---|---|---|
| **iVoice** | 87.5% (G2) | 44.6% (G1) |
| **DrNS** | 53.6% (G1) | 37.1% (G2) |

Table 1: Cell Coverage

Table 1 shows that iVoice users covered a higher percentage of the cells than those using Dragon NaturallySpeaking. In Spreadsheet 1 users who used iVoice covered on average 87.5% of the spreadsheet whereas users who used Dragon NaturallySpeaking only covered 53.6%. For Spreadsheet 2, using iVoice, participants covered on average 44.6% of the spreadsheet whereas participants using Dragon NaturallySpeaking covered 37.1%. Thus, for the first spreadsheet task, the group using iVoice covered more cells than the group using DrNS. For the second spreadsheet task the technologies were reversed to ensure any difference in behaviour could not be attributed to the makeup of the groups. The table shows a clear result with the group who used iVoice again covering more cells.

While the experimental sample is small it is possible to apply statistical methods to investigate whether this difference could be due to random variation. Given the small number of values, twelve in total, non-parametric methods represents the best approach with a rank sum test based on two samples of six values the most natural measure. Based on a one sided test, with an alternative hypothesis that the coverage achieved by users of iVoice exceeds that of users of DrNS, a p-value of 0.045 was determined. This indicates, based on a 5% significance level, that there is sufficient evidence to reject the null hypothesis of no difference in coverage received.

|  | Spreadsheet 1 | Spreadsheet 2 |
|---|---|---|
| **Group 1** | 28.4 | 21.5 |
| **Group 2** | 26.8 | 20.0 |

Table 2: Time spent auditing spreadsheet in minutes

It is important to establish that any difference in performance between the two technologies is not due to a difference in time spent by the groups on the tasks. Table 2 shows the average time each group spent auditing each spreadsheet. It was found that Group 1 spent on average just 1½ minutes more on each spreadsheet than those in Group 2, regardless of the technology that was employed.

|  | Spreadsheet 1 | Spreadsheet 2 |
|---|---|---|
| **iVoice** | 69.4%(G2) | 61.1%(G1) |
| **DrNS** | 63.9%(G1) | 57.4%(G2) |

Table 3: Overall Performance Per Spreadsheet

Table 3 shows the average percentage of errors that were found by each group. It was found that users of iVoice found between 3% and 6% more than users of Dragon NaturallySpeaking. For Spreadsheet 1 those using iVoice found 69.4% of the errors whereas those using Dragon NaturallySpeaking found 63.9%. For Spreadsheet 2 iVoice users found 61.1% of the errors whereas those using Dragon NaturallySpeaking found 57.4%. Following the logic presented above a statistical analysis was performed yielding a p-value of 0.33 which at a 5% significance level indicates that, while there is a difference in average performance there is insufficient evidence to reject the null hypothesis of no difference in performance.

Evaluation of an Intelligent Assistive Technology for Voice Navigation of Spreadsheets
Derek Flood, Kevin Mc Daid, Fergal Mc Caffery, Brian Bishop

In order to determine the effectiveness of each of the features of iVoice a more detailed analysis was conducted. This analysis isolated where each of the iVoice features could have been used, if they were available, and compared these times to where they were used, when participants were using the iVoice technology.

### 6.1 NAVIGATE TO REFERENCED CELLS

The time to navigate between spreadsheets to referenced cells was analysed. To ensure a fair comparison the figures calculated do not include the time to issue the first command which leaves the original cell or the time to leave the referenced cell. The times are measured from when the participants leave the source cell until they enter the destination cell. During the trial this feature was used between 1 and 10 times by each participant. The average values are quoted only for participants who performed this action three or more times.

During the trial it was found that using Dragon NaturallySpeaking users spent on average 4.1 seconds moving from a cell that contains a reference to the worksheet which contained the referenced cell. By using iVoice users can move directly to the referenced worksheet with no time delay.

As well as the four seconds users spent navigating to the referenced worksheet, participants required, on average, a further 2.7 seconds returning to the original cell. This means that through the use of iVoice users can save approximately seven seconds checking one remote reference, as the iVoice system can bring users directly to such a reference in one command and back to the original cell through the "*Jump Back"* command.

The above figures are based on a user moving through at most one intermediate worksheet. As the number of intermediate worksheets increases it is expected that the savings quoted above would also increase.

### 6.2 SCAN COMMAND

The scan function automatically moves through a range of cells, pausing for one second on each cell. This feature proved useful to most participants as they would naturally perform this action while auditing a spreadsheet. This action was performed by participants between 2 and 25 times during the trial.

The aim of the analysis was to concentrate on navigation as much as possible. For this reason the time users spent in the first cell of the range was discarded, as it is believed that users naturally spend longer auditing the first cell in the range and including this would distort the results. The results quoted include the time spent reviewing each cell as it was not possible to isolate the navigation behaviour from the time spent auditing the cell.

Each of the ranges included in the calculation contained a minimum of three cells. It was also determined that, to get a true sense of the time taken, each participant should have scanned at least 3 ranges. For this reason one of the participants was omitted as they failed to scan three valid ranges.

|  | Group 1 | Group 2 |
| --- | --- | --- |
| **iVoice** | 0.93 | 0.97 |
| **DrNS** | 2.62 | 2.92 |

Table 4: Average time (in seconds) per cell in a scanned region

Table 4 shows the average time participants spent on each cell while scanning through a region. It was found that when participants used the iVoice function they were able to spend

Proceedings of EuSpRIG 2008 Conference "In Pursuit of Spreadsheet Excellence " 76
**ISBN : 978-905617-69-2**Copyright: © 2008 European Spreadsheet Risks Interest Group (www.eusprig.org) and Authors

the expected 1 second on each cell in a scanned region. When using Dragon NaturallySpeaking it was found that they spent on average 2.5 – 3.0 seconds on each cell, almost 3 times as long.

### 6.3  JUMP BLANK CELLS

In order to evaluate the effectiveness of the "Jump Direction" command, the time at which users left a cell to the time they entered the next non-blank cell was measured. It was found that participants using Dragon NaturallySpeaking spent approximately 1.3 seconds performing this action. With iVoice the equivalent time would be zero as they are brought directly form one cell to the next non-blank cell. It should also be noted that the time spent in the first cell is not included in these results. While this feature was not used by all participants, a number of participants used it extensively with one participant using it 45 times throughout the trial.

### 6.4  DISCUSSION WITH PARTICIPANTS

Upon completion of the quantitative trial, a structured interview was conducted to find out participants' opinions of the technologies they used and their prior level of experience with both spreadsheet technology and voice recognition technology.

It was found that of the six participants, four of them preferred the iVoice navigation system over Dragon NaturallySpeakings' own navigation system. The participants remarked that the iVoice commands made tasks like moving to remote references easier and also mentioned that it was easier to concentrate on auditing the spreadsheet while using iVoice. One of the participants who expressed no preference between the technologies said that while using iVoice they spent some time reviewing that the coloured cells were the correct references which was distracting them from the debugging task.

When asked, participants remarked that the commands *"Go to Cell"* and "*Move to Cell*" used in Dragon NaturallySpeaking caused them difficulty as the voice recognition engine would often misrecognise parts of the command. Apart from these two commands however participants found the voice recognition software to be accurate.

During the interview participants highlighted a number of issues with the iVoice technology. One such issue relates to the scan function. Participants found that the predetermined wait time was inappropriate, as it was too long for some cells while it was not long enough on others. Another issue that was mentioned was that the scan function would stop one or two cells past the desired cell. This issue is often referred to as overshoot. The reason for this overshoot is that the voice recognition engine takes time to recognise the "*Stop*" command.

### 7  FUTURE WORK

In the future it is hoped that some of the issues identified during the interview could be addressed. By allowing participants to alter the wait time through two additional commands "*Speed Up*" and "*Slow Down*", the scan function could become more efficient and easier to use.

Other enhancements could be made that would partly automate the auditing process. One such enhancement might involve the addition of auditing intelligence to the scan function. If a cell is discovered during the scan procedure that is semantically different than the preceding cells, the scan function could stop automatically on this cell, alerting users to a possible error.

## 8 CONCLUSIONS

In this paper an intelligent navigation system for spreadsheet auditing is proposed. By using the current context of the user, the system, titled iVoice, reduces the number of commands that are needed to perform three common spreadsheet auditing behaviours.

An evaluation of iVoice was conducted through a controlled study which involved both quantitative and qualitative elements designed to address the effectiveness, efficiency and usability of the tool. The study serves as an initial evaluation with the proposal to conduct more extensive trials in the future.

In the study six experienced spreadsheet users were asked to highlight as many errors as they could in two spreadsheets, one with iVoice and the other using a leading voice recognition software, Dragon NaturallySpeaking. Although the trial involved just six participants the cross-over type design allowed the establishment of a statistically significant increase in coverage associated with the iVoice tool. However, where users of iVoice tool outperformed users of the DrNS system the difference could not be shown to be statistically significant.

## 9 ACKNOWLEDGEMENT

This work is supported by the Irish Research Council for Science, Engineering and Technology.

**REFERENCES**

Begel, A. (2004) Spoken Language Support for Software Development *In Proceedings of the 2004 IEEE Symposium on Visual Languages - Human Centric Computing*

Bishop, B., Mc Daid, K. (2007) An Empirical Study of End-User Behaviour in Spreadsheet Debugging *In The 3rd Annual Work-In-Progress Meeting of the Psychology of Programming Interest Group*

Bishop, B., Mc Daid, K., (2008) Spreadsheet Debugging Behaviour of Expert and Novice End-Users *In Workshop on End-User Software Engineering, International Conference on Spreadsheet Engineering (ICSE)*

Burnett, M., Sheretov, A., Ren, B., Rothermel G., (2002) Testing Homogeneous Spreadsheet Grids with the What You See Is What You Test Methodology. *IEEE Transactions on Software Engineering*, 576 -- 594.

Croll, G. (2005) The Importance and Criticality of Spreadsheets in the City of London *In Proceedings of The European Spreadsheet Risk Interest Group Annual Conference*

Fisher, M., Rothermel, G., Brown, D., Cao, M., Cook, C., Burnett, M., (2006) Integrating Automated Test Generation into the WYSIWYT Spreadsheet testing methodology. *ACM Transactions on software Engineering,* 15**,** 150 - 194.

Flood, D., Mc Daid, K., (2007) Voice-controlled Auditing of Spreadsheets *In Proceedings of The European Spreadsheet Risk Interest Group Annual Conference*

Mc Daid, K., Rust, A., Bishop, B., (2008) Test-Driven Development: Can it Work for Spreadsheets? *In Workshop on End-User Software Engineering, International Conference on Spreadsheet Engineering (ICSE), forthcoming,*

Miller, G. (2006) HUD alleges Overpayment for section 8. *Columbia Daily Tribune. February 22nd 2006*

Panko, R., (2005) What We Know About Spreadsheet Errors - Adapted Version (online) available at http://panko.shidler.hawaii.edu/SSR/Mypapers/whatknow.htm [March 14th 2008]

Powell, S. G., Baker, K. R. And Lawson, B. R., (2007) Errors in operational Spreadsheets.

Rust, A., Bishop, B., Mc Daid, K., (2006) Investigating the potential of Test-Driven Development for Spreadsheet Engineering *In Proceedings of The European Spreadsheet Risk Interest Group Annual Conference*

Scansoft (2004) A Dragon NaturallySpeaking Deployment at the Social Security Administration Increased Productivity and Reduced Costs.

Wang, Z., Van De Panne, M., (2006) "Walk to Here" A voice driven Animation System *In Eurographics / ACM SIGGRAPH Symposium on Computer Animation*